\documentclass[
reprint,
superscriptaddress,
amsmath,amssymb,
prb,
]{revtex4-2}

\usepackage{graphicx}
\usepackage{dcolumn}
\usepackage{hyperref}
\usepackage{bm}
\usepackage{esvect}
\usepackage{amsmath}

\usepackage{xcolor}
\begin{document}

\preprint{APS/123-QED}
\title{Measurement-driven neural-network training for integrated magnetic tunnel junction arrays}
\author{William A. Borders}
\email{william.borders@nist.gov}
\affiliation{Physical Measurement Laboratory, National Institute of Standards and Technology, Gaithersburg, MD, USA}%
\author{Advait Madhavan}
\affiliation{Institute for Research in Electronics and Applied Physics, University of Maryland, College Park, MD, USA}%
\affiliation{Associate, Physical Measurement Laboratory, National Institute of Standards and Technology, Gaithersburg, MD, USA}%
\author{Matthew W. Daniels}
\affiliation{Physical Measurement Laboratory, National Institute of Standards and Technology, Gaithersburg, MD, USA}%
\author{Vasileia Georgiou}
\author{Martin Lueker-Boden}
\author{Tiffany S. Santos}
\author{Patrick M. Braganca}
\affiliation{Western Digital Research Center, Western Digital Corporation, San Jose, CA, 95119, USA}%
\author{Mark D. Stiles}
\author{Jabez J. McClelland}
\author{Brian D. Hoskins}
 \email{brian.hoskins@nist.gov}
\affiliation{Physical Measurement Laboratory, National Institute of Standards and Technology, Gaithersburg, MD, USA}%

\date{\today}
             
\begin{abstract}
The increasing scale of neural networks needed to support more complex applications has led to an increasing requirement for area- and energy-efficient hardware. One route to meeting the budget for these applications is to circumvent the von Neumann bottleneck by performing computation in or near memory. However, an inevitability of transferring neural networks onto hardware is the fact that non-idealities such as device-to-device variations or poor device yield impact performance. Methods such as hardware-aware training, where substrate non-idealities are incorporated during network training, are one way to recover performance at the cost of solution generality. In this work, we demonstrate inference on hardware-based neural networks consisting of 20,000 magnetic tunnel junction (MTJ) arrays integrated on a complementary metal-oxide-semiconductor (CMOS) chips in a form that closely resembles scalable and market-ready spin transfer-torque magnetoresistive random access memory (STT-MRAM) technology. Using 36 dies, each containing a MTJ-CMOS crossbar array with its own non-idealities, we show that even a small number of defects in physically mapped networks significantly degrades the performance of networks trained without defects and show that, at the cost of generality, hardware-aware training accounting for specific defects on each die can recover to comparable performance with ideal networks. We then demonstrate a robust training method that extends hardware-aware training to statistics-aware training, producing network weights that perform well on most defective dies regardless of their specific defect locations. When evaluated on the 36 physical dies, statistics-aware trained solutions can achieve a mean misclassification error on the MNIST dataset that differs from the software-baseline by only 2~\%. This statistics-aware training method could be generalized to networks with many layers that are mapped to hardware suited for industry-ready applications.
\end{abstract}

\maketitle

\section{\label{sec:intro}Introduction}
A major drawback of contemporary neural networks is the area and energy cost associated with the performing hardware. For conventional computing systems relying on central and graphical processing units (CPUs and GPUs), numerous cycles of retrieving data from memory and shuttling it to the processor are required even for the most core functions of neural networks~\cite{capraHardwareSoftwareOptimizations2020,canzianiEvaluationNeuralNetwork2017,jouppiDomainspecificArchitectureDeep2018,biancoBenchmarkAnalysisRepresentative2018,xuScalingEdgeInference2018}. For many neural networks, the hardware spends the majority of its time moving data back-and-forth~\cite{coatesDeepLearningCOTS2013,siuMemoryRequirementsConvolutional2018,moolchandaniAcceleratingCNNInference2021}, a crippling factor for conventional systems known as the von Neumann bottleneck.  Furthermore, natural language processors such as GPT-4 using transformer-style models show a trend of doubling in scale every two months, a rate 24 times faster than the performance scaling of silicon hardware predicted by Moore's law~\cite{AICompute, mehonicBraininspiredComputingNeeds2022}. While efforts to improve GPU performance have reduced this scaling gap, the von Neumann bottleneck---as well as the amount of memory area required to store larger network information---remains a difficult barrier for embedded applications such as self-driving vehicles and drones~\cite{arnautovicEvaluationArtificialNeural2021,vitaLowPowerHWAcceleratorAI2020,nicosiaEfficientLightHarvesting2018}. This has led to research in neuromorphic devices, circuits, and systems that can combine with CMOS to match the growing complexity.

One approach to improving neural network hardware is in-memory computing~\cite{sebastianMemoryDevicesApplications2020}, an architecture that reduces the von Neuman bottleneck's effect by performing computation in memory. The intrinsic crossbar array structure of conventional memory is highly suited to perform vector-matrix-multiplication (VMM), a key process in neural networks representing the network layer output as the multiplication of the layer input vector and a weight matrix connecting neurons of adjacent layers. In a crossbar array structure, this corresponds to applying a vector of voltages to the rows of the memory array, and measuring the resultant currents on the columns. Approaches to in-memory computing using conventional CMOS structures such as static~\cite{JMLR:v18:16-456,biswasCONVSRAMEnergyEfficientSRAM2019} and dynamic random access memory~\cite{gaoComputeDRAMInMemoryCompute2019,yooLogicCompatible4T2019, wangInmemoryComputingArchitecture2021} show mitigation of the von Neumann bottleneck, thereby achieving lower area and energy consumption. Demonstrations using non-volatile resistive memory technologies potentially further improve efficiency~\cite{xiaoAnalogArchitecturesNeural2020,legallo64coreMixedsignalInmemory2023a}. 

Recent research on neural networks shows that the low classification error of fully-connected networks using high-bit precision can still be maintained on network models with sparse weight connections~\cite{hanDeepCompressionCompressing2016,hoskinsStreamingBatchEigenupdates2019} or drastically reduced weight precision~\cite{trusovFastImplementation4bit2021,sunUltraLowPrecision4bit2020,wuRotationConsistentMargin2020}; both of these strategies reduce the resources required for computation. Even binary neural networks (BNNs)~\cite{courbariauxBinarizedNeuralNetworks2016}, where the activation of neurons and/or the synaptic weights are represented by single bit values, have shown equivalent performance compared to the above methods on small datasets, with a potential for application to larger networks~\cite{qinBinaryNeuralNetworks2020}. 

A promising candidate for combining in-memory computing and BNNs is an array of back-end-of-the-line-compatible magnetic tunnel junctions (MTJs) that can each represent two stable resistive states using the tunnel magnetoresistance (TMR)~\cite{julliereTunnelingFerromagneticFilms1975} effect. As the core component of magnetic random access memory (MRAM)~\cite{kentNewSpinMagnetic2015,ikedaPerpendicularanisotropyCoFeBMgO2010}, MTJs are a prime choice for representing weights in a BNN, reaching embedded memory sizes greater than 16~Mb\cite{leeWorldmostEnergyefficientMRAM2022}, boasting high endurance, and potentially replacing conventional memories for embedded applications\cite{worledgeSTTMRAMStatusOutlook2022a}. Recent works with passive~\cite{goodwillImplementationBinaryNeural2022,zhouExperimentalDemonstrationNeuromorphic2021} or CMOS-integrated~\cite{jungMRAMInmemoryComputing2022,jungCrossbarArrayMagnetoresistive2022} crossbar arrays of MTJs experimentally demonstrate BNN inference on various datasets and show performance comparable to software baselines.

For many hardware accelerator applications, performing inference on the hardware first involves training a weight matrix solution offline and then downloading trained weight values onto the hardware. In the case of an MTJ-based BNN, a device programmed to the high resistance state might represent an offline weight value of zero. However, an inevitability of using physical hardware in the analog domain is the potential for non-idealities in devices or supporting hardware. Non-idealities such as line resistance~\cite{goodwillImplementationBinaryNeural2022,jeongParasiticEffectAnalysis2018}, device variations~\cite{xiaoDeviceVariationAwareAdaptive2022b, gokmenAccelerationDeepNeural2016,gonugondlaVariationTolerantInMemoryMachine2018}, bit errors~\cite{hirtzlinImplementingBinarizedNeural2019,boniardiPhysicalOriginResistance2011,buschjagerBitErrorTolerance2021}, and poor overall device yield can result in imperfect manifestations of the offline weight matrix. Even in conventional CMOS technology, bit faults require methods such as burn-in testing or error correction codes in order to ensure consistent operation. Several stopgaps exist to combat these non-idealities, such as hardware-aware~\cite{raschHardwareawareTrainingLargescale2023} in-situ training of the network where network error is sampled directly from the hardware~\cite{yaoFullyHardwareimplementedMemristor2020,gonugondlaVariationTolerantInMemoryMachine2018}, or ex-situ training where statistical models add noise to the weight values~\cite{doevenspeckNoiseTolerantTernary2021,laubeufDynamicQuantizationRange2022}. Experiments and simulations have shown that arrays of MTJs can recover classification accuracy by implementing hardware-aware training to compensate for device-to-device variations~\cite{xiaoDeviceVariationAwareAdaptive2022b} and bit errors~\cite{hirtzlinImplementingBinarizedNeural2019}. 

In this work, we study offline training methods on non-ideal arrays by fabricating 36 dies each containing 20,000 MTJs integrated with CMOS into a crossbar array structure, where each die represents the weights of a two-layer BNN. These arrays are described in Sec.~\ref{sec:design} and the network structure is described in Sec.~\ref{sec:soft}. Section~\ref{sec:single} reports that ex-situ training of each die where the unique defect characteristics of each die are considered recovers mean classification accuracy to within 2~\% of the software baseline.

One drawback to training around the \emph{unique} defects of a unit of hardware, however, is loss of generality of the trained weight matrix. In practice, industrial manufacturers may produce millions of chips that could have non-idealities in different locations, and these defects may even change during the device's lifetime. Therefore in Sec.~\ref{sec:robust} we present an extension of hardware-aware training, which we refer to as statistics-aware training, that compensates for one of the most impactful of these non-idealities---namely, shorted devices---regardless of the location of these devices. To accomplish this, we train not according to the usual loss function of the software neural network but rather with a loss function that averages over the defect statistics observed in hardware. We realize an efficient training method over this defect-statistics-aware loss function via a double-batch outer product training method where many parallel instances of the same network have their weights in the first network layer randomly chosen to be defective. The gradient averaged over all instances is then used to update a single statistics-aware weight solution. We use this method to produce 100 solutions which are tested across all dies and find that the mean difference in classification error from the software baseline is 2~\%, demonstrating a robust substrate-agnostic solution. The value to which defective weights are set during training not only determines the software baseline classification error, but also controls the agreement between hardware and the performance variation between dies. Investigating the sensitivity of the hardware network output with respect to each weight shows the output is overall less sensitive to changes in weights for a statistics-aware network when compared with non-robust methods. This observation leads to discussion regarding the balance between optimal layer size and defect density in hardware neural networks.

\section{\label{sec:results}Experimental setup}
\subsection{\label{sec:design}Design and test of 20,000 MTJs integrated with CMOS}
\begin{figure*}
        \centering
        \includegraphics[width=0.8\textwidth]{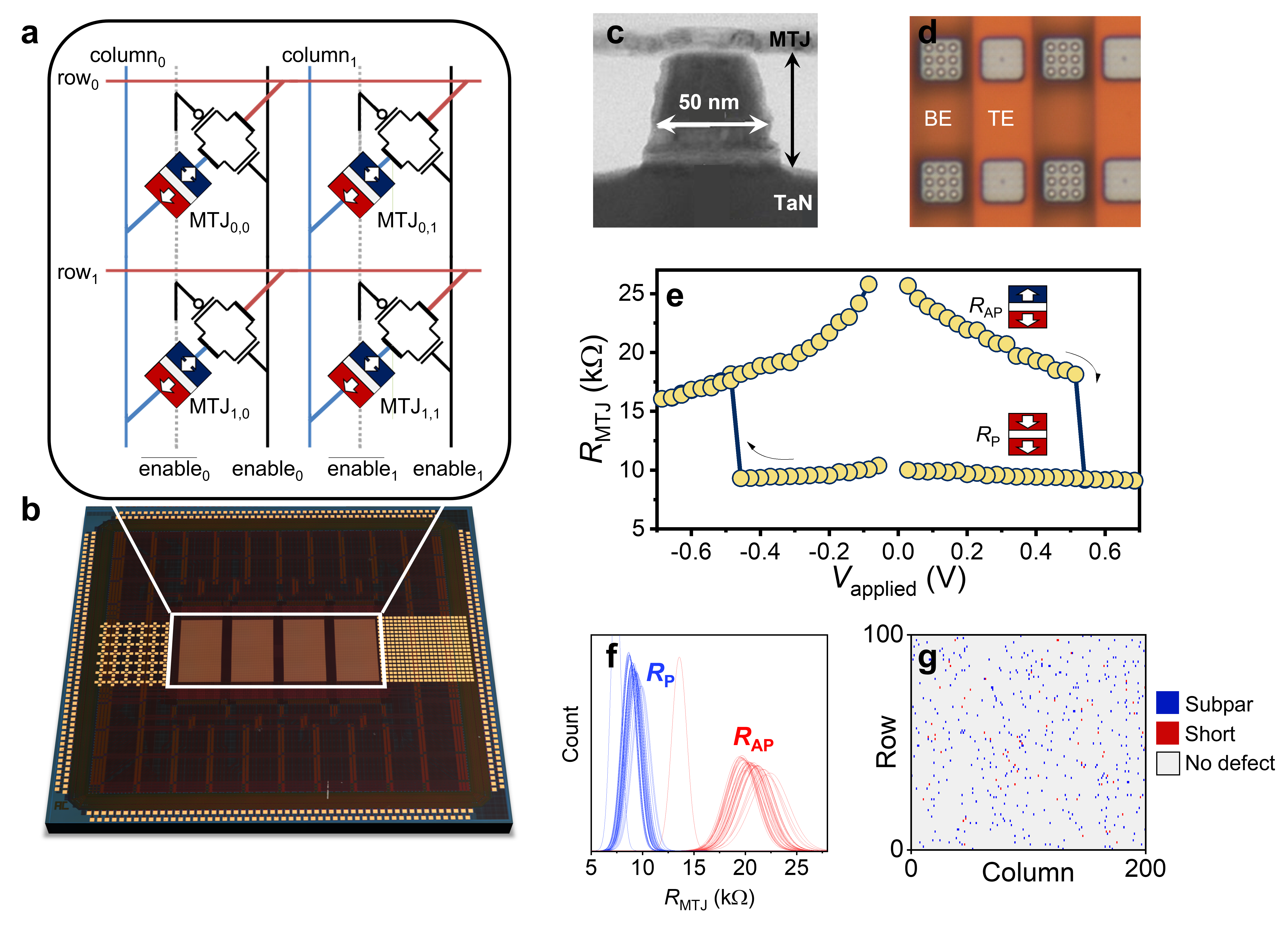}\par
        \caption{Fabrication and characterization of a CMOS-integrated array of 20,000 MTJs. (a) Schematic of a 2 $\times$ 2 portion of the 2T-1R crossbar array. Voltages are applied to the columns or rows while selecting a transistor column with enable lines. (b) Optical microscope image of one die containing 20,000 MTJs integrated with 40,000 transistors. 403 metallized pads for contact with probecard needles surround the array. (c) Transmission electron microscope cross-sectional image of an MTJ pillar patterned above TaN pads in contact with tungsten vias. (d) Optical microscope view of top and bottom electrodes for MTJ integration. The bottom electrode (BE) is patterned with a 3 x 3 array of MTJ vias while the top electrode (TE) is patterned with a single MTJ. Patterning the MTJ via array is an efficient process that removes extra steps during fabrication. Furthermore, the diameter of each via is two orders of magnitude larger than the MTJ pillar, producing only a trivial contribution to the measured resistance of the single MTJ.  (e) Representative resistance vs. voltage curve for the MTJs used in this work. (f) Gaussian fits to the histograms for filtered \(R_\text{P}\) and \(R_\text{AP}\) values of MTJs in each die. (g) One example of the defect locations within a single die. Shorted devices display a constant resistance between 100~$\Omega$ and 1~k$\Omega$, while subpar MTJs switch, but with resistances ranging between 1~k$\Omega$ and 12~k$\Omega$.}
        \label{fig:design of 20k}
\end{figure*}

    We design and fabricate a 2-transistor-1-resistor (2T-1R) test vehicle capable of housing 20,000 two-terminal devices in a crossbar array architecture shown in Figs.~\ref{fig:design of 20k}(a,b). The entire chip consists of 100 rows $\times$ 200 columns which are partitioned into four subarrays of 50 columns each. A two-bit digital signal selects the subarray, allowing for characterization of 5000 devices at a time. The two-transistor design acts as a pass gate circuit with an n-channel metal-oxide semiconductor (NMOS) and p-channel metal-oxide semiconductor (PMOS) transistor in parallel; a digital enable signal determines whether the pass-gate operates in a digital or analog mode. 

    The choice of a 2T-1R structure as opposed to a 1T-1R or passive array structure was made to produce a general test vehicle for characterizing many two-terminal novel device technologies. When the device properties call for current compliance, the PMOS can be turned off and the NMOS can be used. For all experiments in this work, measurements were performed in digital mode: the gate voltage on the NMOS (PMOS) transistor is power (ground), reducing as much as possible any series resistance with the MTJ. For ideal MTJs with narrow property distributions, a passive array is the best option for the highest density on chip. In a practical situation where device properties are well characterized but non-idealities such as sneak paths and line resistance exist, a final product of this chip would use a 1T-1R structure to isolate sections of the chip for measurements. The 2T-1R initial designs for the 180-nm CMOS are fabricated in a commercial foundry without the MTJs and received with the final metal layer removed, exposing the tungsten (W) vias underneath. 

    Heterogenous back-end integration begins by patterning TaN pads above the vias to control the interfacial roughness on which the MTJ films are grown. These films are deposited using magnetron sputtering and patterned into 50~nm nominal diameter pillars and 1~{\textmu}m diameter via arrays on alternating TaN pads by electron beam lithography and Ar ion milling. Fig.~\ref{fig:design of 20k}(c) shows a transmission electron microscope cross-sectional image of a single MTJ pillar in contact with the TaN pads. A final metallization process is performed to connect the top electrode of the MTJ to the row lines and the bottom electrode to the transistor circuit [Fig.~\ref{fig:design of 20k}(d)]. The MTJ patterning process is repeated 36 times across 36 dies of a 150 millimeter silicon wafer. 
    
    Each array is accessed through 403 pads which connect to the row, column, transistor gate, power, and ground lines of the crossbar. A probecard interfacing with an offboard 5-channel source-measure-unit and switch-matrix uses 403 tungsten needles to contact the pads, one die at a time. The probecard powers the die with 3.3 V DC, which is used by corresponding enable lines to supply voltage to the NMOS and PMOS transistors. Port-to-port measurements are performed by applying a voltage on either the rows or columns and reading current from the opposite terminal.

    The MTJs used in this work consist of two ferromagnetic layers, a fixed and a free layer, separated by an insulating tunneling barrier. The magnetization of the fixed layer is pinned by a synthetic antiferromagnet and the energetically stable orientation for the free layer is in either a parallel (P) or anti-parallel (AP) configuration relative to the magnetization of the fixed layer. The MTJ configuration can be switched by the application of a current that induces a spin-transfer torque~\cite{SLONCZEWSKI1996L1,PhysRevB.54.9353,nmat3311Brataas} on the magnetization of the free layer. MTJs are characterized by their tunnel magnetoresistance (TMR), where the TMR of the MTJ shows a low (high) resistance for the P (AP) configuration, and the TMR ratio is defined as 
    \begin{equation}
        \text{TMR ratio} = \frac{R_{\text{AP}}-R_\text{P}}{R_\text{P}}. 
    \end{equation}
    Fig.~\ref{fig:design of 20k}(e) shows a representative magnetization reversal curve of the MTJs used in this work. The curve is obtained by sweeping the applied voltage from 0~V to 0.75~V and back to 0~V in the positive direction, followed by the same process in the negative direction. The MTJ shows clear switching from the AP state to the P state around 0.5~V and from P to AP around $-0.5$~V. When characterizing all 20,000 MTJs on each die, we use a single write voltage of 0.75~V to ensure switching of each device.
    
     We apply a screening test to each die of 20,000 MTJs to produce distributions of \(R_\text{P}\), \(R_\text{AP}\), and TMR ratio. Those distributions then determine the yield and locations of defective devices. MTJs first receive a write voltage of $-0.75$~V to switch the configuration to the AP state, after which a small 100~mV signal is applied to probe the resistance of the device. The same process is applied with the same magnitude but opposite polarity to write the configuration to the P state. Cycling through all 20,000 devices in the array requires approximately two minutes. We determine \(R_\text{P}\) and \(R_\text{AP}\) from the measured voltage and current during the read steps. The yield for each die is determined by the number of devices with \(R_\text{P}\geq 6~\text{k}\Omega\), \(R_\text{AP}\leq 30~\text{k}\Omega\), and TMR ratio $\geq 0.6$. Across all 36 dies the median yield is 99.2~\% with a standard deviation of 0.65~\%. Fig.~\ref{fig:design of 20k}(f) shows a Gaussian fit to the histograms of \(R_\text{P}\) and \(R_\text{AP}\) after filtering out defective devices. The distribution for one die exhibits a considerable offset from the trend shown by the majority of the dies, but seeing as this offset occurs in both distributions for \(R_\text{P}\) and \(R_\text{AP}\), we expect the effect on neural network performance to be trivial. In general, the variation of the mean of each distribution is less of a factor compared to the width of each separate distribution. 

    After the screening process, the type of defect is separated by severity, where two types of defects exist in the die. Most defects consist of MTJ pillars that are electrically shorted, exhibiting resistances between 100~$\Omega$ and 1~k$\Omega$. The other type of defect consists of 'subpar' devices that show magnetization reversal but exhibit resistance ranges between 1 k$\Omega$ and 12 k$\Omega$. A defect map for a single die is shown in Fig.~\ref{fig:design of 20k}(g) with shorted and subpar devices highlighted. Across all dies, defects are randomly distributed with only three dies containing large areas of defective devices due to non-idealities introduced during fabrication.

\subsection{\label{sec:soft}Neural network architecture design and mapping onto MTJ hardware}
    To test the performance of the crossbar array as a neural network, we use the MNIST dataset\cite{deng2012mnist}, which includes 60,000 training and 10,000 test inputs. Each input is a $28 \times 28$-pixel grayscale image of a handwritten digit associated with an output target represented by a binary array of size 10. Each pixel contains a value ranging from 0 (black) to 1 (white) proportional to that pixel's intensity. We construct a simple two-layer feed-forward network, shown in Fig.~\ref{fig:network to hardware and pure software results}(a), that can be mapped onto our hardware array. 
    
    Implementing VMM with 784-pixel input images on hardware would require more resources than our hardware supports, so each input image is transformed  by scaling and cropping into a $10 \times 10$ pixel array. The complete two-layer neural network architecture includes 100 input neurons, 90 hidden neurons, and 10 output neurons---one for each handwritten digit---which all together require a $100\times 90$ weight matrix for the first layer and a $90 \times 10$ weight matrix for the output layer. Weights in the network are represented using two MTJs, where the weight is proportional to the conductivity difference between the excitatory MTJ\(_\text{e}\) and inhibitory MTJ\(_\text{i}\). Under this representation, network weights can take on a ternary value of either $-1$, 0, or 1. Excitatory and inhibitory MTJs are connected in the same row but adjacent columns; when VMM is performed, layer outputs are calculated by subtracting the two column currents. The complete mapping of the network architecture onto the crossbar array is shown in Fig.~\ref{fig:network to hardware and pure software results}(b). 
    
    Network training is performed offline, using backpropagation and stochastic gradient descent to minimize the cross-entropy loss of the network. During training, the forward and backward passes are performed using a ternarization of an underlying collection of real-valued continuous weights. After each weight update, the real-valued weights are ternarized according to a threshold which maps values to $-1$, 0, or 1. The network is trained to convergence for 50 epochs after which the final ternary weights and biases are saved to be used in the crossbar. We repeat this training process 100 times, each with a unique random weight initialization to produce 100 potential configurations of trained weights; each configuration is defined as a weight matrix \emph{solution} of the training problem.

    Under the current experimental setup, we are not able to implement VMM in parallel on the network, which would require 100 unique voltages sourced from 100 unique source-measure unit channels. Work in Ref.~\cite{goodwillImplementationBinaryNeural2022} shows that if the devices are operated in a linear regime, a full VMM can be represented by applying a read voltage element-by-element to each device and summing the resulting currents in software. We have verified that the applied voltages for this work are in this regime and our measurements satisfy the superposition principle of linear circuits. For this reason, the following results are labeled as hardware emulation, because the conductances of MTJs are used in software to determine the output currents. Furthermore, due to the sheer volume of programming steps for 100 solutions across 36 dies, we take advantage of the very low cycle-to-cycle variation  in MTJ resistance (see Appendix~\ref{app:yield}) and program solutions onto the weight matrices in software using measured \(R_\text{P}\) and \(R_\text{AP}\) values obtained during screening of the dies.
    
    Inference classification error of the hardware emulation is determined by first encoding the ternary weights during training onto the MTJ array in software by converting weights into pairs of MTJ conductances. We map ternary matrix weights into conductance pairs such that $[\text{AP, P}] \cong -1$, $[\text{AP, AP}] \cong 0$, and $[\text{P, AP}] \cong 1$.  VMM is emulated by transforming the 100 pixel inputs into input voltages through multiplication by a read voltage \(V_\text{read} = 100\)~mV and then multiplying those voltages with the read conductance of each measured MTJ. Currents calculated for each column are normalized into dimensionless quantities by dividing out both \(V_\text{read}\) and a conductance-valued hyperparameter \(G_\text{norm}\). Mathematically, the VMM operation to obtain the currents for each layer is
    \begin{align}
        \vec{y}_\text{outputs} &= \frac{\vec{I}_\text{columns}}{V_\text{read}G_\text{norm}}\\
        &= \frac{\hat G_e - \hat G_i}{G_\text{norm}}\cdot\vec{x}_\text{inputs},
    \end{align}
    where \(\vec{x}\) and \(\vec{y}\) are dimensionless inputs and outputs, \(\hat{G}_\text{e}\) and \(\hat{G}_\text{i}\) are the respective excitatory and inhibitory weight matrices. A bias is applied to the 90 outputs for layer 1 (\(z_0\), ..., \(z_{89}\)) and passed through a hyperbolic tangent (tanh) activation function to determine layer 2 inputs (\(a_0\), ..., \(a_{89}\)). Layer 2 outputs are determined in an identical manner to layer 1. During training, outputs are fed through a softmax activation function for the target output to determine the loss function. In the inference stage, the argument of the maximum layer 2 output is compared to the target. A tally of incorrect classifications is incremented for any prediction that does not match the target and a total classification error is defined as the ratio of incorrect classifications divided by the size of the dataset. 

\section{Training the neural network\label{sec:training}}
In the following sections, we discuss and compare multiple ways of training offline solutions for the MTJ crossbar, referring to three types of solutions: ``defect-free'' solutions, ``hardware-aware'' solutions, and ``statistics-aware'' solutions. The first is the result of training the network architecture in the conventional way with no hardware-informed considerations; the second includes details of a \emph{particular manifestation} of a crossbar into the training method; the last uses only \emph{general statistics} of the hardware to create solutions that are more universally applicable regardless of specific defect configurations found on a single die. Before reporting our results on using these three solution methods in hardware, we briefly outline the theoretical relationships between these methods. Furthermore, we refer to the classification error of the solution trained offline as the ``software-baseline'' and the same solution's performance on hardware as the ``hardware emulation''.

In the training of a typical software neural network, an unknown true loss function $\ell(x;\mathbf{W})$ associated with the learning task exists which is a function of the network weights $\mathbf{W}$ and any particular input $x$ to the neural network. The goal is to minimize the total expected loss 
\begin{equation}
    L(\mathbf{W}) = \int \ell(x;\mathbf{W})\,\mathcal{D}x\label{eq:true-loss}
\end{equation}
over the real-valued parameter space for $\mathbf{W}$, with measure $\mathcal{D}x = g(x)\,dx$ carrying some probability density $g(x)$ that encodes the likelihood of encountering any particular network input $x$. Although in practice we cannot access the true loss function, any training data set $X$ of size $N_X$ allows us to sample $L$ at a finite number of points, which gives rise to the \emph{empirical} loss function 
\begin{align}
    \tilde L(\mathbf{W}) = \frac{1}{N_X}\sum_{j=1}^{N_X} \ell(x_j;\mathbf{W})
    \label{eq:empirical-loss}
\end{align}
which is what is actually minimized by stochastic gradient descent during training. Here the distribution $g(x)$ is automatically encoded through the assumption that $X$ is a statistically representative sample of inputs. Solutions obtained by minimizing $\tilde L$ via stochastic gradient descent are what we refer to as defect-free solutions.

How do hardware-aware training methods fit into this framework? When a specific defect configuration is chosen---that is, when the unique details of defects in a single die are specified---we can imagine encoding those details as part of the function realized by the neural network. In this case, the loss function $\ell$ must be modified according to those defects, realized in practice by manifesting the effect of defects during the forward pass of a neural network training routine: we replace $f(x;\mathbf{W})$ with $f(x;\xi(\mathbf{W}))$ in Eq.~\eqref{eq:true-loss}, where $\xi$ is a function that applies defects to the nominally programmed weights $\mathbf{W}$. Realizing the particular details of $\xi$ can be crucial for offline training; as such, chip-in-the-loop~\cite{tam1990learning} and physics-aware training~\cite{wrightDeepPhysicalNeural2022} methods sometimes use the hardware itself in the forward pass to ensure that $\xi$ is accurately captured. 

This work aims to find solutions that perform robustly \emph{regardless of which defective hardware they are used on}. We want a solution to minimize error over an ensemble of possible realizations of the defects. To express this sensitivity of training to the defect statistics, we construct a new loss function, 
\begin{align}
    L^\text{HW}(\mathbf{W}) = \int \ell(x;\xi(\mathbf{W}))\,\mathcal{D}\xi\,\mathcal{D}x,\label{eq:distributional-loss}
\end{align}
which now contains an extra integration over possible defect modes $\xi$. Given a nominal set of weights $\mathbf{W}$, $\xi(\mathbf{W})$ returns a random but plausible set of effective parameters realized on physical hardware, with statistical measure $\mathcal{D}\xi = f(\xi)\,d\xi$ informed by measured statistics of the actual hardware. 

The primary type of $\xi$ considered in this work is the one that randomly maps the elements of $\mathbf{W}$ to the saturation value representing an electrical short with some probability, the choice of which emerges in Eq.~\eqref{eq:distributional-loss} via a strongly restricted probability density $f(\xi)$. It is also easy to imagine including device-to-device variation, stuck weights, open connections, or other defect types under the same framework of an expected value over $\xi$.

The remainder of this section is structured as follows. In Sec.~\ref{sec:single}, we investigate hardware-aware methods that use details of single-crossbar defects to optimize performance on a specific die. Then in Sec.~\ref{sec:robust}, we consider a practical implementation of the stochastic gradient descent over the statistics-aware loss function $L^\text{HW}$ described above. Because Eqs.~\eqref{eq:true-loss} and \eqref{eq:distributional-loss} do not coincide in general, it is expected that robust statistics-aware solutions will perform worse on defect-free inference tasks than traditional solutions would -- and vice versa. Sec.~\ref{sec:robust} shows that this hypothesis is vindicated.

\subsection{\label{sec:single}Improving classification error with hardware-aware training methods}
\begin{figure*}[hbt!]
    \centering
    \includegraphics[width=0.9\textwidth]{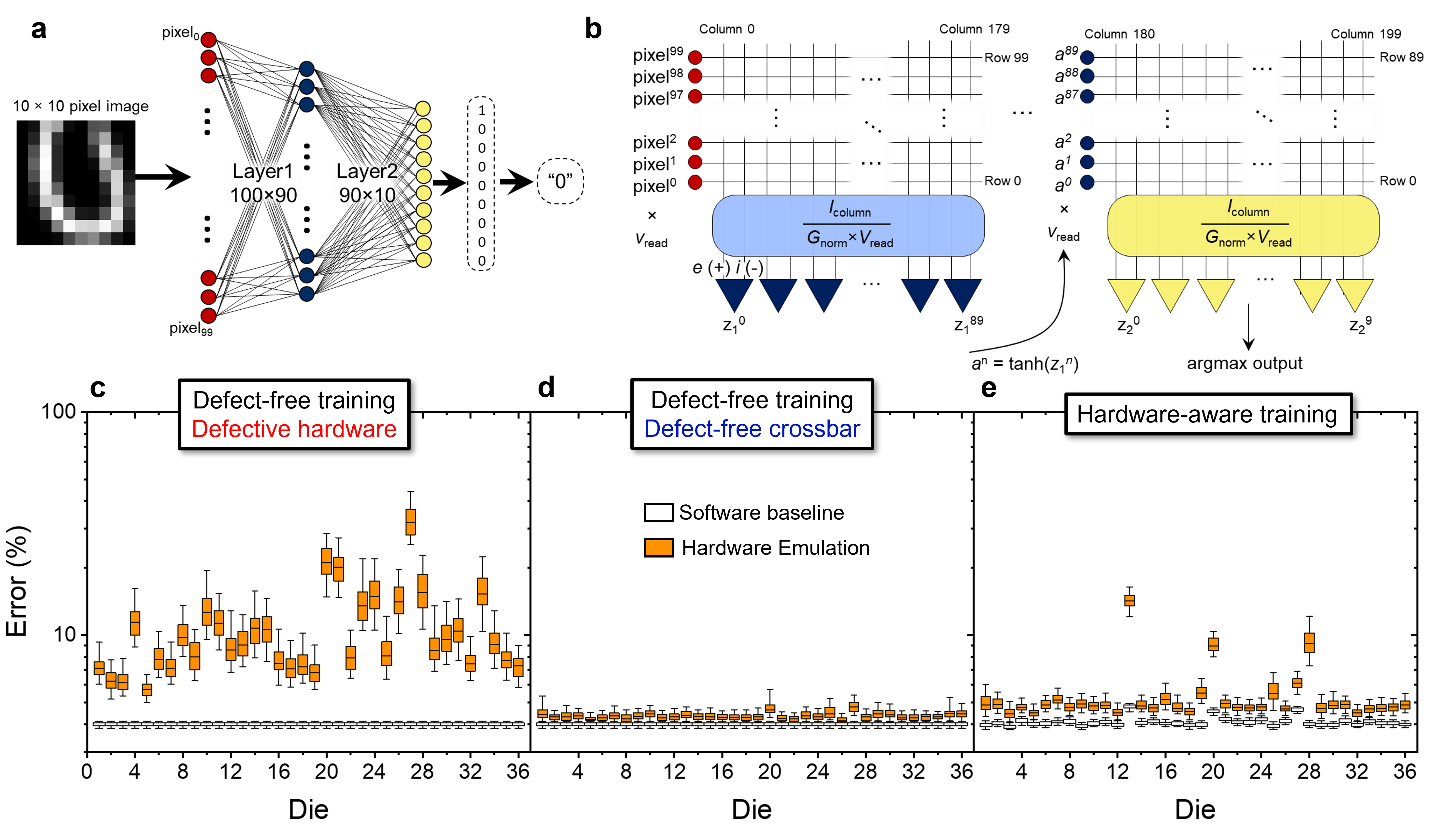}\par
    \caption{Mapping of neural network to crossbar array hardware and inference performance on the reduced MNIST dataset. (a) Visual representation of the network architecture used for inference. Images are scaled and cropped from 28 $\times$ 28 to 10 $\times$ 10 pixel images and input to a two-layer feed-forward network. Each neuron in the output layer represents one possible handwritten digit. (b) Schematic of the neural network mapping to the MTJ crossbar array. The hardware-equivalent neural network function is the same color as the corresponding function shown in (a). Each weight is determined by the difference in conductance of two MTJs in adjacent columns and the complete network utilizes 19,800 of the 20,000 MTJs. (c) Box and whisker plot showing inference classification error of the MNIST test dataset on 36 different crossbar array dies using 100 defect-free weight matrix solutions. White boxes represent the classification error software baseline(defect-free solution) and colored boxes represent the same solutions emulated on the MTJ hardware. The horizontal line within each box indicates median error and the box and whiskers represent the 25th to 75th and 5th to 95th percentiles, respectively, for 100 unique weight matrix solutions. (d) Classification error of the defect-free solutions in (c) when defective MTJs are replaced with the mean ON and OFF MTJ conductance of the die. (e) Classification error for 100 hardware-aware solutions for each die where each solution is trained around the unique defects of each die.}
    \label{fig:network to hardware and pure software results}
\end{figure*}

In this section, we first present hardware emulation results of performing inference on 100 solutions across all 36 dies when all defective devices are included. We then show how an ideal crossbar without defects would perform, followed by description of a hardware-aware training method that compensates for the defects of specific die to improve performance. The classification error on the MNIST test dataset for each die is plotted as a box and whisker plot in Fig~\ref{fig:network to hardware and pure software results}(c). White boxes represent the distribution of the software baseline, with a mean error of 4~\%, and colored boxes show the distribution of error for the hardware emulation on each die. The software baseline is the same for each die and repeated to easily compare performance. There is a large variation between classification errors; some dies perform well with less than 10~\% error while some show a higher error and variation. It is interesting to note that these large errors occur despite the average device yield of 99.2 \%.

In the following results, we address three metrics for performance: mean error, variation of solutions on a single die, and variation of mean error between dies. The large variation of classification error within each die can be attributed to the locations of defects. Due to the random initialization of weights, the converged network is sensitive to values of weights in certain locations. For a portion of the 100 solutions, the locations of these sensitive weights will overlap with defect locations, adversely affecting classification error. On the other hand, the large variation between dies is caused by a variation in the number of defects. 

To compare the performance of a defect-free solution on defective hardware with the performance of a defect-free solution on an ideal crossbar, defective locations in each die are replaced with conductance values equal to the mean AP and P conductance of the die (Fig.~\ref{fig:network to hardware and pure software results}(d)). The ideal crossbar shows a mean error of 4.3~\%, only a 0.3~\% difference with the software-baseline. The discrepancy between the software-baseline and those of the hardware emulation is due to the variation of the ON and OFF weight state as a result of device variation. We perform a simulation of networks with increasing values of variation and show that variations similar to the physical hardware match the increase in error (see Appendix~\ref{app:dtdvariation}).

To bridge the gap between these results, 100 solutions for each die are trained while considering the properties of the underlying substrate. During training, weights at locations where defects exist are set to a constant value equivalent to the weight the defect's resistance would induce. This can produce weights ranging from 10 to 80 for shorted devices, and 1 to 10 for subpar devices. Figure~\ref{fig:network to hardware and pure software results}(e) shows the result of performing hardware-aware training for each die. Compared to Fig.~\ref{fig:network to hardware and pure software results}(c),(d) the baseline software error varies between dies. This effect arises from a different number of defects per die. The mean classification error across all dies is 5.2~\%, a value only 0.9~\% larger than the result for ideal crossbars, demonstrating that hardware-aware training produces performance comparable to defect-free solutions. 

Several significantly poorer-performing outliers are seen in Fig.~\ref{fig:network to hardware and pure software results}(e), in particular the ones associated with dies 13, 20, and 28. These large deviations can be traced to large particles introduced to the wafer during fabrication, resulting in large areas of defective devices or entire rows of devices shorted. Since the origin of these non-idealities is significantly different from random distributions of shorts and subpar devices, the results of these three dies are excluded from mean performance. 

Further discrepancy between the defect-free solutions and the hardware emulations is in part due to device variation inducing variation of weights around -1, 0, or 1. In addition to device variation, an incomplete representation of the defective device's weight could increase error. Because each weight is represented by two MTJs, it is impossible to know what state the non-defective device of the pair should be during training. The impact is trivial for shorted devices, but could be comparable for certain subpar devices.

During hardware-aware training we observe a crippling increase in the classification error if defects are included in layer 2. It is unclear if the increase is because of the ratio of defects to the size of the layer or whether it depends on the location of defects in the previous layer. For this work, we focus our investigation on the impact of defects in a single layer, replacing layer 2 defects with working devices that remain in the same column. We justify this method by observing that MTJ crossbar arrays are best used for layers in the network where large volumes of devices/weights are needed to gain the most benefit from area and energy costs. Final layers are typically smaller when compared to the rest of the network and due to their sensitivity to defects, could be implemented in defect-free CMOS. In future work, we plan to investigate more complicated networks with more hidden layers, and the effect that defects in each layer has on classification error. 

\subsection{\label{sec:robust}Training and validating statistics-aware solutions}
One observation of the hardware-aware training method is that solutions trained for one die do not produce the same performance on other dies. In any practical application where millions of dies are manufactured, training each die would require too many resources. Furthermore, devices could start to fail if exposed to electromagnetic fields, radiation, extreme temperatures, or more commonly, electrostatic discharge. These factors are an even larger concern in environments where all the above issues commonly exist, for example space-based applications. In this section we describe a robust statistics-aware training method that performs similarly across all dies, regardless of defect configuration, by attempting to minimize the statistics-aware loss function of Eq.~\eqref{eq:distributional-loss}.

\begin{figure*}
    \centering
    \includegraphics[width=\textwidth]{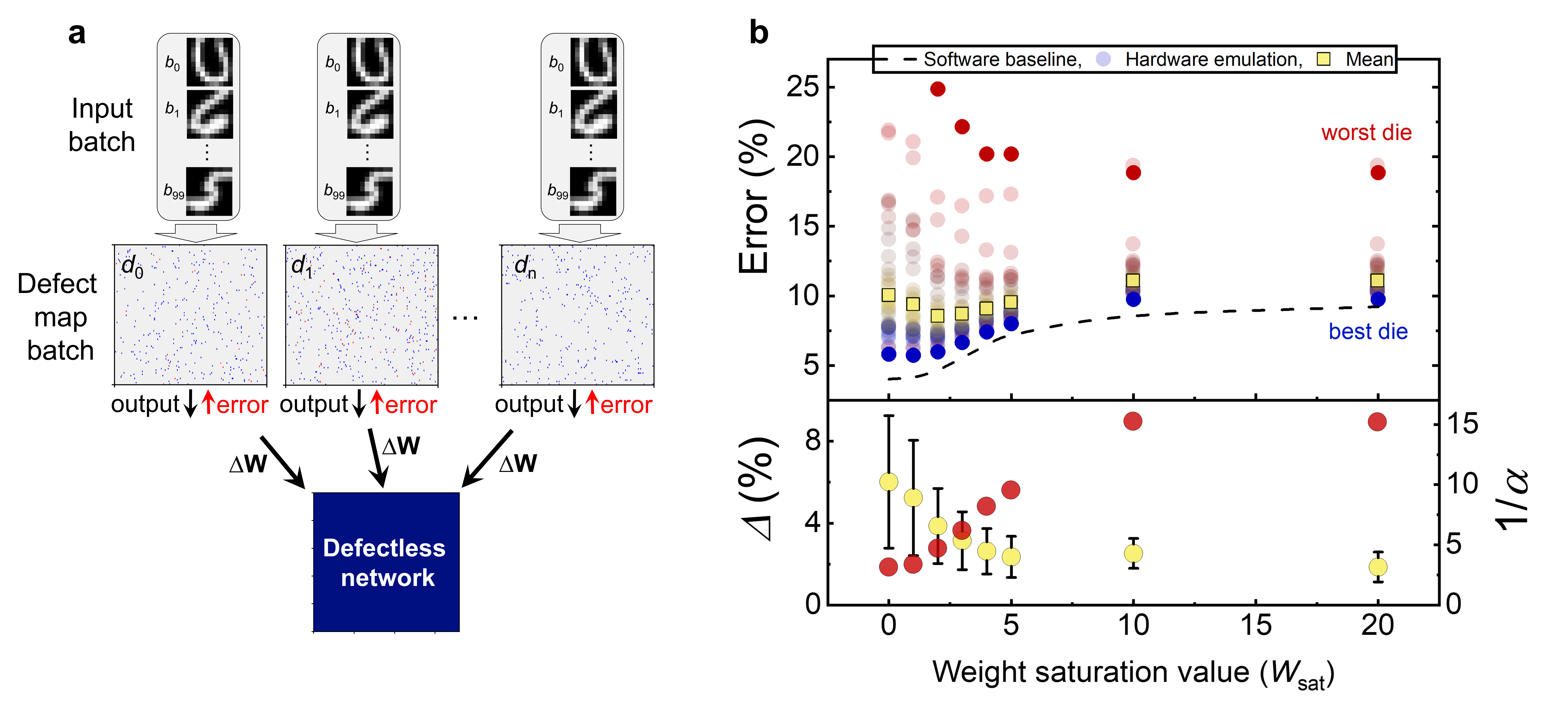}\par
    \caption{Training of statistics-aware solutions. (a) Visual representation of training a statistics-aware solution. For each batch of input images, an identical size of defect maps are randomly generated. In this work, both the input image batch size and the defect map batch size is 100. (b)(top) Classification error of all 36 dies for $W_{\text{sat}}$ ranging from 0 to 20. Each circle represents the mean error of 100 solutions for one die assuming a single $W_{\text{sat}}$. Yellow squares represent the mean across all dies, while the dotted line represents the mean software-baseline classification error. Bold dots signify the best (blue) and worst (red) performing dies. (bottom) \textit{$\Delta$}, defined as the difference in error between the software-baseline and the emulated hardware (yellow circles), and 1/$\alpha$, defined as the inverse of the coefficient of variation for hardware emulation (red circles). Error bars represent the one standard deviation width of the distribution of 36 dies. Both values are determined by ignoring the outliers shown in the above plot.}
    \label{fig:Robust solution results}
\end{figure*}

For a single solution to work on the exponentially large number of possible defect configurations, the method of training the solution must account for the variability in defect location. This suggests that networks with a small subset of weights dominating network performance are also susceptible to increased error in the presence of defects. A change in any of these important weights has a large impact on network performance. In order to train robust solutions, the statistical properties of the defective hardware must be accounted for in the training process. 

In conventional stochastic gradient descent training methods a batch (sometimes called a minibatch) of possible inputs are randomly chosen and used to construct an estimator of the empirical loss Eq.~\eqref{eq:empirical-loss}. That is, we construct the gradient of
\begin{equation}
    \tilde{L}_B(\mathbf{W}) = \frac{1}{N_B}\sum_{j=1}^{N_B}\ell(x_j;\mathbf{W})
\end{equation}
where $B\subseteq X$, of size $N_B$, is our randomly sampled batch from the full training dataset. The weights are then subjected to an update rule $\mathbf{W}\leftarrow\mathbf{W}-\eta\nabla_\mathbf{W} \tilde{L}_B$ for some learning rate $\eta$. This is minibatch gradient descent, the overwhelmingly standard basic algorithm for training deep neural networks.

Since the goal is to minimize Eq.~\eqref{eq:distributional-loss}, we extend the above notion of a batch to also include a randomly sampled set of defect maps $\Xi$. Statistics-aware solutions are trained by producing a total batch given through the outer product of the usual training batch $B$ with our defect map batch $\Xi$, the latter sampled from a representative distribution of possible defect locations (Fig.~\ref{fig:Robust solution results}(a)) informed by measured statistics of the experimental hardware. In this way, training proceeds by moving along the gradient of the statistics-aware minibatch loss
\begin{equation}
    \tilde{L}^\text{HW}_{B,\Xi}(\mathbf{W}) = \frac{1}{N_B}\frac{1}{N_\Xi}\sum_{i=1}^{N_\Xi}\sum_{j=1}^{N_B} \ell(x_j;\xi_i(\mathbf{W})),
\end{equation}
which can be used to update the weights in exactly the same way as stochastic gradient descent. Although not investigated here, we note that since all our defect statistics are captured within the loss function itself, one could nominally subject the statistics-aware loss to any modern stochastic-gradient-descent optimizer such as Adam~\cite{kingma2017adam} or AdaGrad~\cite{Duchi_Hazan_Singer_2011}. We leave questions of how defect statistics interact with such algorithms to future work.

The training process begins from a pristine network without defects and with weights initialized with the normally-distributed Xavier initialization~\cite{glorot2010understanding}. The weights of this network after training represent the final solution to be mapped onto the 36 dies for testing inference. Defect maps are generated first by selecting a number $n$ of total defects sampled from a normal distribution fit to the experimental characterization; the $n$ locations of these defects are sampled uniformly at random over all possible sites on the crossbar. In constructing the defect maps, all defective weights are assumed to take on the same large saturation value $W_{\text{sat}}$ corresponding to the large conductance of a shorted device, though it is easy to imagine a more complex algorithm that samples that value as well. 

Because performance depends on the saturation value $W_\text{sat}$, we conduct studies of 100 different statistics-aware solutions for $W_\text{sat}$ ranging from 0 to 20. Since the defects are encoded in the loss, both forward and backward passes encode defect map information over the batch of input images and their accumulated gradients are used to update the real-valued weights of the network. With every new batch of training data, a new batch of defect maps is generated as well. Since each weight update operation involves a batch of images and defects, we refer to this method as a double-batch outer product update. 

The performance of the statistics-aware solution is shown in Fig.~\ref{fig:Robust solution results}(b). For each $W_{\text{sat}}$ we plot the mean classification error for 100 solutions on each die as well as the overall mean classification error. With a larger $W_{\text{sat}}$, the classification error for both the software baseline and the hardware emulation converges to 10~\%. At the same time, the variation in performance between dies improves at higher $W_{\text{sat}}$. Looking closer at the performance of each die, some dies perform well even at low $W_{\text{sat}}$ and proceed to perform worse when trained with stronger defect maps, while poorer performing dies show less difference in error with the software-baseline. This behavior implies that for any given defect configuration with a given number of shorted and subpar devices, there is an optimal $W_{\text{sat}}$. 

Determining an optimal value for this system would not be likely to apply to other hardware and network sizes. Instead, we focus on the trends of performance -- in particular, the difference in classification error between hardware emulation and software baselines, and the variation of performance between dies. The lower plot of Fig.~\ref{fig:Robust solution results}(b) shows that at higher $W_{\text{sat}}$ the mean difference between the hardware emulation and the software-baseline converges to between 0.01 and 0.02 and the variation between these differences shrinks. This improvement in variation between dies comes at the cost of a higher overall software baseline classification error. This is depicted in the bottom panel of Fig.~\ref{fig:Robust solution results}(b), which shows that the inverse of the coefficient of the variation ($\alpha$) of the mean error on all dies increases. In a similar fashion, the variation in classification error between solutions on a single die also shrinks at higher $W_{\text{sat}}$ (see Appendix~\ref{app:robust}). If a hardware application requires lower overall error and removal of poorly performing dies is acceptable, then a lower $W_{\text{sat}}$ may be preferable. If variation between dies is a concern but some performance loss is tolerable, then a larger $W_{\text{sat}}$ is preferred. 

\subsection{Analysis of network sensitivity}
To clarify the underlying factors determining the robust solution performance, we investigate the sensitivity of the network to weight variations. We define the sensitivity $\mathcal{I}$ of the network to changes in each weight as
\begin{equation}
    \mathcal{I}(w_{ij}) = \left\langle\left(\frac{\partial \ell(x;\mathbf{W})}{\partial w_{ij}}\right)^2\right\rangle_x
\label{eq:FI}
\end{equation}
where \(w_{ij}\) is the weight located in column $i$ and row $j$, $\mathbf{W}$ encodes the network weights, $x$ represents the an individual input to the network, \(\ell (x;\textbf{W})\) represents the loss function of the network on that input $x$, and the angle brackets indicate the expected value averaging over $x$ at fixed $\textbf{W}$. 

For both the defect-free and the statistics-aware solution, there are two quantities to compute to characterize the statistics-aware training process. 
The first is calculating the sensitivity,  $\mathcal{I}$, when evaluated with the defect-free loss function. The second is to evaluate both solutions in the presence of defects. To do this, we extend $\mathcal{I}$ to include defect-map averaging in a way similar to what was done when passing from Eq.~\eqref{eq:true-loss} to Eq.~\eqref{eq:distributional-loss}, giving a defect-statistics-aware sensitivity
\begin{equation}
    \mathcal{I}^\text{HW}(w_{ij}) = \left\langle\left(\frac{\partial \ell(x;\xi(\mathbf{W}))}{\partial w_{ij}}\right)^2\right\rangle_{x,\xi}
\label{eq:FI-defects}
\end{equation}
that correctly captures the geometry of the statistics-aware loss function used to train the statistics-aware solution. We consider both of these approaches to computing sensitivity below.
To estimate $\mathcal{I}^\text{HW}$, the trained solutions are copied 100 times and for each copy defects with $W_{\text{sat}} = 20$ are randomly dispersed throughout their first layers. We average the mean squared gradients over the 100 defective networks. This process is repeated for each of the 100 solutions. We compute the sensitivity to each of the $100 \times 90$ weights in the first layer and accumulate their statistics.

\begin{figure}
    \centering
    \includegraphics[width=0.95\linewidth]{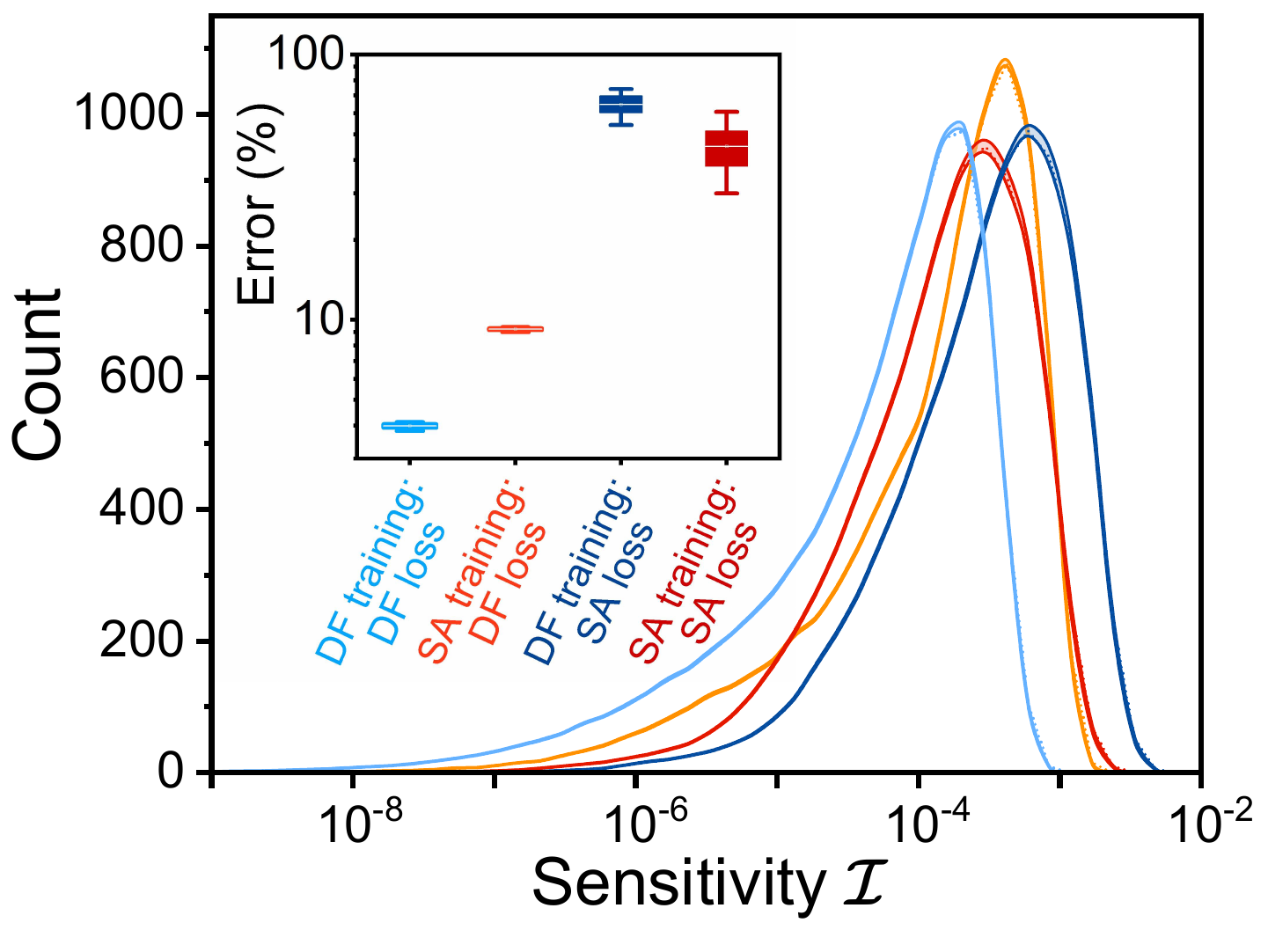}\par
    \caption{Histograms of $\mathcal{I}$ and $\mathcal{I}^\text{HW}$ among the layer 1 weights across 100 pure software solutions and 100 statistics-aware solutions evaluated on the training dataset. Light blue and orange curves represent the defect-free (DF) and statistics-aware (SA) solutions, respectively, determined with $L$ (defect-free loss). Navy blue and red curves represent the defect-free and statistics-aware solutions, respectively, determined from $L^\text{HW}$ (statistics-aware loss).  Each curve represents the mean of 100 solutions for each bin in the histogram and the uncertainty in the mean at each value. The uncertainty in the mean for each curve is negligble, causing the mean curve and the error bands to overlap each other. All histograms are binned in an identical manner. (inset) resulting classification error for the four histograms shown, plotted on a logarithmic scale. Boxes and whiskers represent the 25th to 75th and 5th to 95th percentiles, respectively. It is important to note the high classification error for the solutions evaluated on the statistics-aware loss is the effect of choosing a value of 20 for $W_{\text{sat}}$. Actual dies contain a distribution of defect values and thus show improved performance.}
    \label{fig:FI Dist}
\end{figure}

Fig.~\ref{fig:FI Dist} shows the histograms for $\mathcal{I}$ and $\mathcal{I}^\text{HW}$ of weights in the first layer calculated using the method described above. Four curves are shown: an software-baseline trained solution without defects (defect-free loss), an defect-free solution with defects (statistics-aware loss), a statistics-aware solution without defects (defect-free loss), and a statistics-aware solution with defects (statistics-aware loss). Each curve represents a mean sensitivity for each weight averaged over 100 different solutions and error bands denote the uncertainty in the mean. The inset shows a box-and-whisker plot of the resultant classification error for the same set of solutions. The software-baseline trained solutions perform better than the statistics-aware trained solutions when evaluated with no defects, but perform worse when evaluated in the presence of defects. 

The computed sensitivities are consistent with the overall performance.
For the case where the sensitivity $\mathcal{I}$ of both the defect-free and statistics-aware solutions are calculated on $L$, the defect-free solution case exhibits a lower overall sensitivity and error. The correlation between sensitivity and error is expected on the grounds that stochastic gradient descent is known to seek out generalizable solutions, that is, solutions in flat basins of the loss function~\cite{doi:10.1073/pnas.2015617118}. 
By contrast, the same two sets of solutions swap their relative error and sensitivities when evaluated in the presence of defects. In the case of a statistics-aware loss where defects are present during training, the solution trained to be statistically aware shows a lower mean $\mathcal{I}^\text{HW}$ compared to the defect-free case.

\begin{figure}[hbt!]
    \centering
    \includegraphics[width=\linewidth]{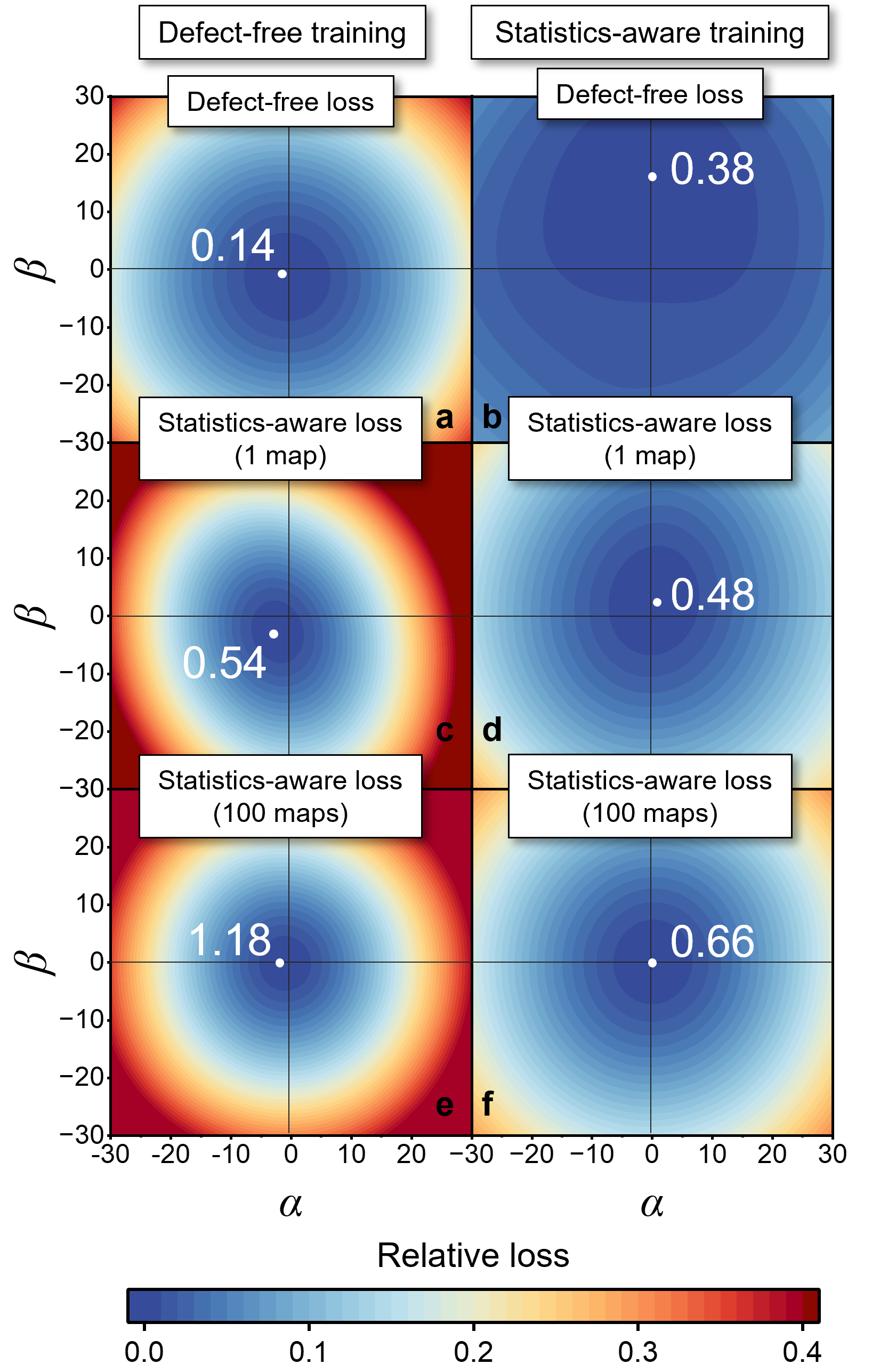}\par
    \caption{Relative loss landscapes for defect-free and statistics-aware training cases, evaluated on the training dataset. (a,b) The relative loss landscape calculated for a defect-free trained solution (a) and a statistics-aware trained solution (b) evaluated on the defect-free loss, $L$. The relative loss is defined as the difference between the loss at each point and the absolute minimum loss obtained in each case. (c,d) The  relative loss landscape for the defect-free and statistics-aware solutions, evaluated in the presence of defects, $L^\text{HW}$. (c) and (d) only consider the effects of 1 unique defect map. (e,f) The  relative loss landscape for the defect-free and statistics-aware solutions, evaluated in the presence of defects, plotted as an average across 100 randomly generated defect maps. White dots represent the location of the absolute minimum and labels represent the value at the minimum. The same two random directions $\delta$ and $\eta$ defined in Eq.~\eqref{eq:Loss land eq} are used for each of the six plots. Within each training method, we also evaluate the loss landscape on an identical weight solution.}
    \label{fig:loss landscape}
\end{figure}
The differences between curves within a single evaluation scheme (i.e. $\mathcal{I}$ or $\mathcal{I}^\text{HW}$) can be understood when considering that the sensitivity function describes the loss curvature of the network with respect to the weights. The converged weights in either training scheme are close to a minimum of the defect-free loss function, typically in a flat basin~\cite{doi:10.1073/pnas.2015617118}. This results in low overall sensitivity since the gradient of the loss function is very small. The fact that solutions trained to minimize $L$ and $L^\text{HW}$ each show superior generalizability under evaluations of $\mathcal{I}$ and $\mathcal{I}^\text{HW}$ respectively is an expected result in that sense.

We can gain a more intuitive understanding of how the evaluation of $\mathcal{I}^\text{HW}$ differs from $\mathcal{I}$ by visualizing the calculation of $\mathcal{I}^\text{HW}$ as a statistical averaging of $\mathcal{I}$ over a cloud of different weight configurations $\{\xi_j(\mathbf{W})\}_j$ centered about $\mathbf{W}$. During the statistics-aware double batch training, the loss is minimized over this point-cloud, not over the single point $\mathbf{W}$. This implies that whatever region in weight space that best minimizes a single point (the minimization of $L$) may not be the same region that minimizes this cloud-averaged loss (the minimization of $L^\text{HW}$). 

If stochastic gradient descent has indeed found a (comparatively) generalizable solution for $L^\text{HW}$, then we would expect the local curvature to be (comparatively) flat at many points in the defect map ensemble. But there is no reason to expect this in the minimization of $L$; evaluating $\mathcal{I}^\text{HW}$ on the pure software solution means sampling apparently random loss function derivatives far from the $\mathbf{W}$ in a fashion that the network was never exposed to during the training process, leading to a higher $\mathcal{I}^\text{HW}$ for the software weights.

To gain a qualitative understanding of the loss around the minima for each case listed above, we perform calculations to visualize the loss landscape. We do so by choosing two randomly determined directions in the parameter space of the first-layer weights, scaling them, and shifting the parameters in the solution by the scaled amount ~\cite{li2018visualizing}. We define a new configuration $\theta$ at which the loss is calculated as
\begin{equation}
    \theta =  \theta^\text{*}+\alpha\frac{\delta}{\lVert\delta\rVert}+\beta\frac{\eta}{\lVert\eta\rVert}
\label{eq:Loss land eq}
\end{equation}
where $\theta^\text{*}$ represents the initial parameter configuration, $\alpha$ and $\beta$ are both scalar values ranging from $-30$ to 30, and $\delta$ and $\eta$ are two randomly chosen unit vectors in parameter space. Unit vectors are generated by sampling a Gaussian random variable with mean 0 and standard deviation 1. For each combination of the scalar values $\alpha$ and $\beta$, we assess the loss of the network and repeat this process for each case studied in Fig.~\ref{fig:FI Dist}. An important note to make here is that the loss landscapes are generated by only applying $\delta$ and $\eta$ to layer 1. We justify this choice based on the fact that this we focus on the impact that strongly defective devices in a single layer have on the output of the network. Thus the objective of viewing the loss landscape is to investigate the impact defects have on the local characteristics of the landscape.

Fig.~\ref{fig:loss landscape}(a) and (b) plot the defect-free and statistics-aware trained solutions, respectively, evaluated with $L$. Two main differences can be seen. One is the apparent flatness of the statistics-aware training method compared to the defect-free training. The second is the shift of the absolute minimum away from the initial parameters $\theta^*$. Both observations are expected: training a network with the statistics-aware method will on average create a flatter landscape due to the statistics of many defect maps used to train the network. A shift implies that evaluating the statistics-aware solution on an defect-free loss is a poor match, simply because the solution is trained with large saturated weights. In contrast, the defect-free and statistics-aware solution evaluated on $L^\text{HW}$ show opposite characteristics. Compared to the defect-free loss, an defect-free trained solution evaluated in the presence of defects shows an order of magnitude higher minimum and a significant shift away from the origin (Fig.~\ref{fig:loss landscape}(c)). Conversely the statistics-aware solution in the presence of defects shown in Fig.~\ref{fig:loss landscape}(d) retains most of its flatness with a minimum very close to the origin. 

The two plots shown in Fig.~\ref{fig:loss landscape}(c),(d) are evaluated on a single defect map to show the agreement of the training method and the underlying hardware when considering a single defective die. We also investigate the loss landscape averaged over 100 randomly generated defect maps and plot them in Fig.~\ref{fig:loss landscape}(e) and (f). While the difference in minima and flatness remain, the minima appear nearly at the origin. This is simply an artefact of averaging many unique landscapes with minimum close to the origin. The key observation is that for 100 different defect maps, the statistics-aware solution will produce a flatter and overall lower magnitude loss landscape, demonstrating a solution that is less sensitive to weight perturbations.

\section{Summary and Outlook}
In this work, 36 dies each containing a crossbar array of 20,000 CMOS-integrated MTJs are used to characterize the performance of a hardware emulation of a binary neural network on the MNIST dataset. We investigate the effect of strongly defective devices on the hardware's classification error in a context where large variation in performance between dies is observed. By implementing a hardware-aware training method that compensates for defective device locations by strengthening the weights in other information pathways, we show that the error for each die can recover to levels comparable to that of an ideal, defect-less crossbar. 

We venture beyond hardware-aware training of particular dies toward statistics-aware training for the entire population of dies, implementing a robust training method that accumulates the network gradient of many defective networks onto a single solution to efficiently represent the parameter space. Statistics-aware solutions where the defective weights are saturated to large values during training show performance closest to software baselines, at the cost of increased misclassification rates. One key observation is that the variation between dies at high $W_{\text{sat}}$ is an order of magnitude smaller than for smaller values. Investigating the local geometry of the loss function for weights in the first network layer shows that, in statistics-aware solutions, the overall sensitivity of weights to perturbations is reduced compared to the defect-free case. This suggests that a single solution trained with this method is less sensitive to the location of defects than a na\"ively trained software solution and can produce reliable and robust performance across multiple different dies.

One factor to consider regarding these results is how they impact hardware at the scale of practical applications.  The two-layer network we investigated here is one of the simplest implementations for neural networks; in practice, deep neural networks contain many layers with varying sizes, architectures, and importance. Investigating the performance impact of defects across multiple layers and establishing the optimal ratio of defect density to layer size---possibly as a network-specific hyperparameter---remain open questions. 

Regarding the proposed double-batch training algorithm, more questions of parameterization remain. Of immediate importance is identifying optimal batch sizes for defect maps. A longer term theoretical question is whether there exists a critical capacity for the number of defect maps ``learned'' by a single layer. Unlike training data, which is highly structured and generally contains most of its significant information in a small (compared to the input dimensionality) number of singular vectors, the defect maps we consider are highly random and are expected to have comparatively flat spectra. This change in the information geometry of the ``data'' presented to the network could have far reaching theoretical consequences for the training algorithms used to guide and optimize stochastic gradient descent for statistics-aware loss functions. Refinements of our existing algorithm and the adaptation of more complex training algorithms to the statistics-aware loss may have the potential to improve training performance beyond what has been presented here.

This experimental work uses MTJs to represent weight values, but the observations made on hardware-aware training, statistics-aware solutions, and weight sensitivity may translate to any crossbar array consisting of resistive devices. While MTJs carry the benefit of fast write speeds, low operating voltages, high endurance, and market-readiness, their ON/OFF ratios are much smaller compared to other resistive memory technologies. A lower ON/OFF ratio translates to a smaller maximum size of the crossbar array, ultimately determined by the line resistance of the substrate. 

As technology progresses, defective devices in hardware-based neural networks will remain an inevitability. This work suggests that at least for binary systems, there are methods for training networks to improve the fidelity of the hardware to comparable levels with software, even in the presence of non-perturbative defect modes, taking full advantage of the significant area and energy savings of in-memory computing schemes over software-based von Neumann approaches.
\section{ACKNOWLEDGEMENTS}
WAB and AM contributed equally to this work. We thank Thomas Boone for fruitful discussions. This work was funded by the National Institute of Standards and Technology. AM acknowledges support under the Cooperative Research Agreement Award No. 70NANB14H209 through the University of Maryland and NSF grant number CCF-CISE-ANR-FET-2121957.
\appendix
\section{Device-to-device variation impact on classification error}\label{app:dtdvariation}
We investigate the effect that device-to-device variation has on the classification error of the hardware emulation. In the main text, it is stated that the discrepancy between the performance of the hardware emulation and the defect-free solution for the hardware-aware training method is due to device-to-device variations inducing a distribution around the ternary weights. To show this, we simulate the MTJ substrate having various levels of variation in resistance. For each simulation, a crossbar is generated by sampling normal distributions with mean equal to the mean P and AP resistance states of the experimentally measured hardware and standard deviation determined by the coefficient of variation, defined as \(\sigma=\alpha\times\mu\), where $\sigma$ is the standard deviation, $\alpha$ is the coefficient of variation, and $\mu$ is the mean. This generated crossbar is then used in the same manner as the other results presented in this paper.
\begin{figure}[hbpt!]
    \centering
    \includegraphics[width=0.9\linewidth]{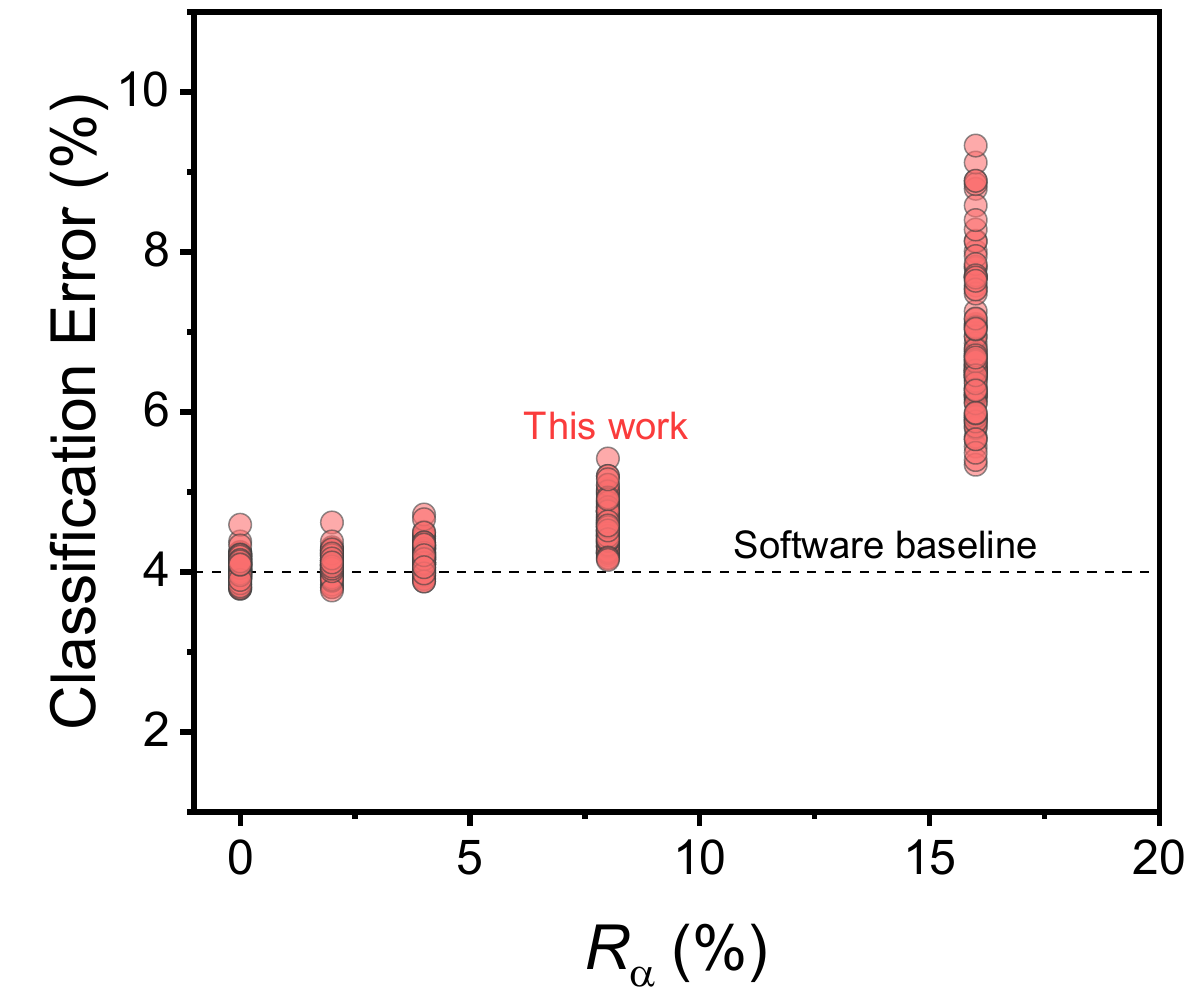}\par
    \caption{Simulation results of inference classification error on the test data set using MTJ resistance values sampled from a distribution with increasing variation. Each point in the plot shows the results of inference for a defect-free solution when tested on hardware with a certain coefficient of variation in resistance.}
    \label{fig:variation vs. error}
\end{figure}

Fig.~\ref{fig:variation vs. error} shows the classification error of the test dataset for 100 pure software solutions (without defects) each on a crossbar with increasing levels of variation. As expected, at 0~\% variation, the error does not differ from the software performance since there are only two values that the MTJs can represent. However as the variation increases, the discrepancy with the defect-free performance begins to diverge. It is worth noting that at 8~\% variation, the discrepancy of the simulated hardware with defect-free performance is of the same level as that of the hardware-aware results shown in Fig.~\ref{fig:network to hardware and pure software results}(e).

\section{Variation of statistics-aware solution performance within each die}\label{app:robust}
\begin{figure}[hbpt!]
    \centering
    \includegraphics[width=0.8\linewidth]{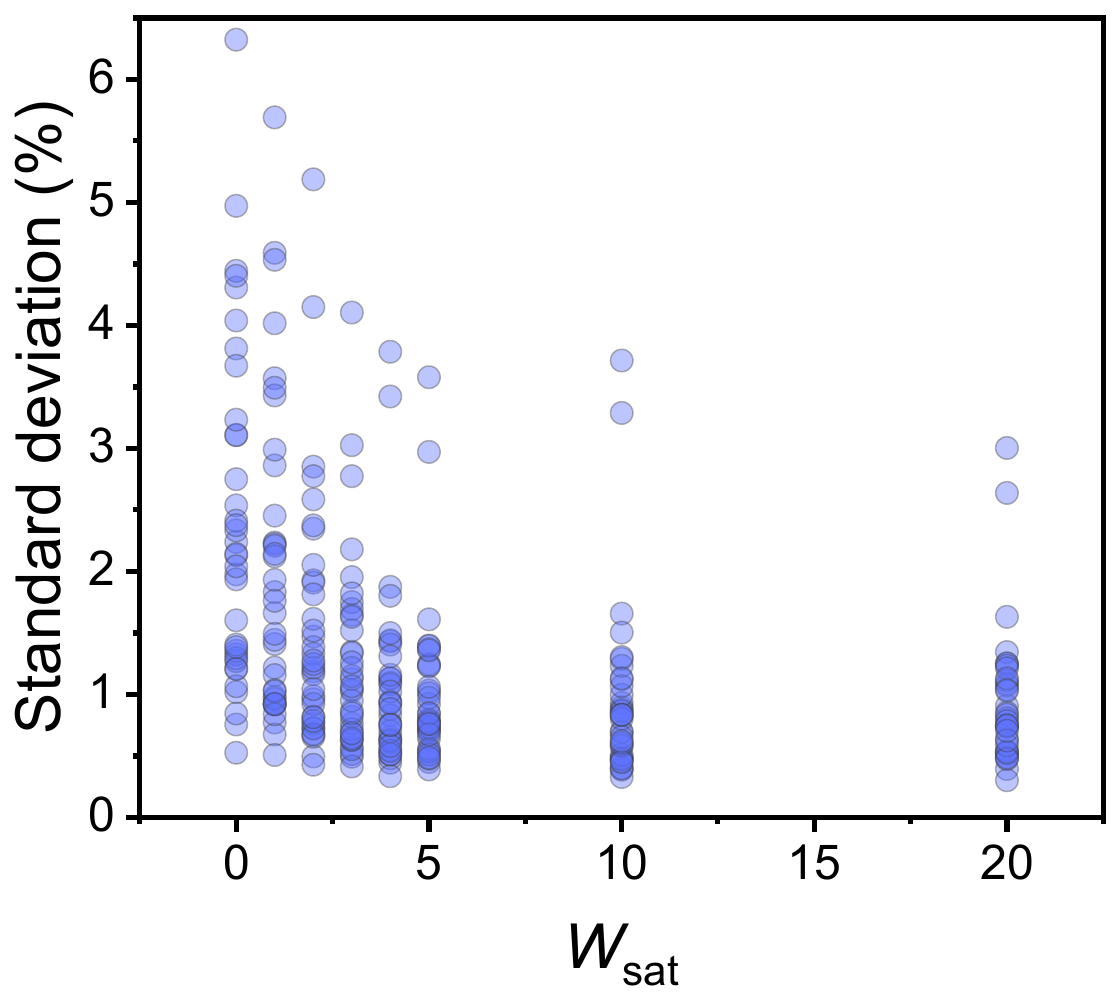}\par
    \caption{Standard deviation of classification error of each die, plotted as a function of $W_{\text{sat}}$.}
    \label{fig:variation of robust}
\end{figure}

Here we discuss the variation of the statistics-aware solution performance within each die. In Fig.~\ref{fig:Robust solution results}(b) the classification error for statistics-aware solutions are plotted as a function of $W_{\text{sat}}$. Each dot plotted represents the mean error of that die gathered from 100 uniquely trained solutions. Fig.~\ref{fig:variation of robust} plots the standard deviation of the 100 solutions for each die. Similar to the trend shown before where the variation between dies shrinks, the variation of error within each die converges to less than 2~\% (ignoring outliers). This trend of reducing variability between solutions confirms the validity of the trend seen in Fig.~\ref{fig:Robust solution results}(b), demonstrating that the statistics-aware training method does in fact reduce the variability of performance between dies.
\section{Yield statistics of all dies and cycle-to-cycle variation of MTJ resistance}\label{app:yield}
In this appendix, we first show cycle-to-cycle variability of the resistance of each MTJ within a single die followed by a plot of heat maps showing the device yield of each die. As mentioned in the main text, inference is performed by emulating the 36 experimentally measured dies. Performing inference on each solution uses identical values for resistance, where in a completely hardware implementation, resistance values will vary. To investigate the levels of resistance variation of the dies in this work, we perform 100 cycles where each MTJ is passed through the screening test mentioned in \ref{sec:design}. 
\begin{figure}[hbpt!]
    \centering
    \includegraphics[width=0.8\linewidth]{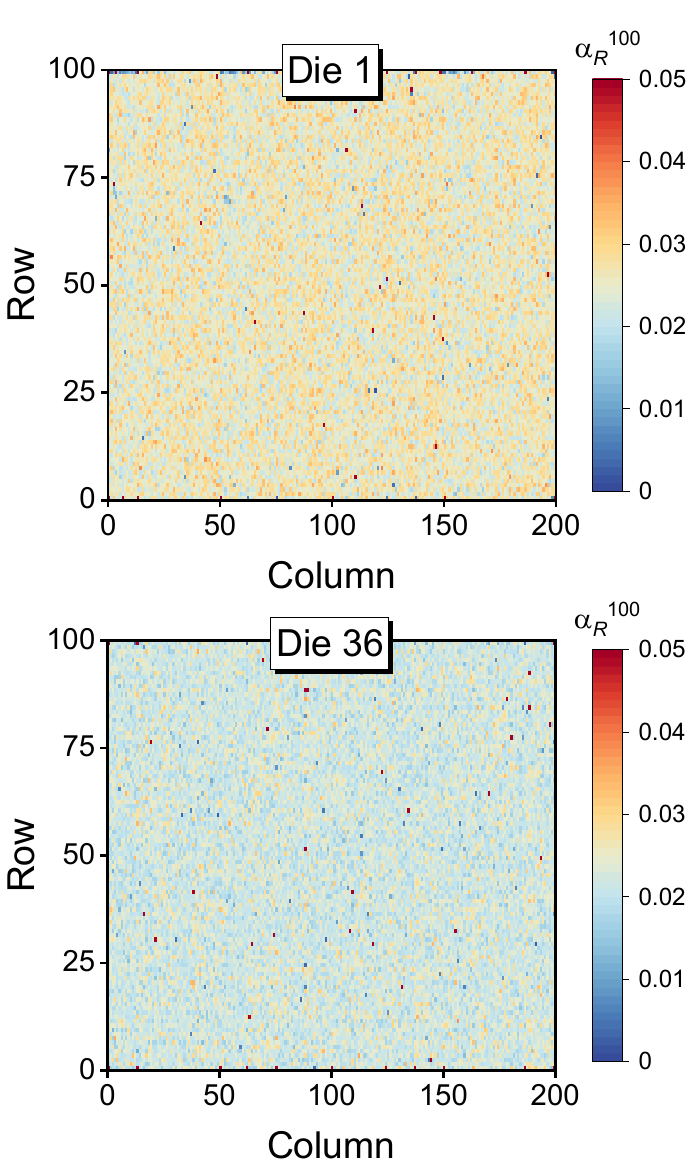}\par
    \caption{Coefficient of variation for each MTJ gathered from 100 cycles of the screening process described in the main text. Results for Die 1 and Die 36 are shown to demonstrate the uniformity across the entire wafer. At these levels of variation, non-defective device's P-state resistance varies by only 200~$\Omega$. This resistance change is at the same scale of the electrically shorted devices, producing the seemingly larger varying devices in shown in red.}
    \label{fig:cycle-cylce var}
\end{figure}
Fig.~\ref{fig:cycle-cylce var} plots the coefficient of variation (defined as the standard deviation divided by the mean) for the P-state resistance \(\alpha_R\) plotted for each of the 20,000 MTJs. This cycle-to-cycle variation test is performed for Die 1 and Die 36, radially on opposite ends of the wafer, to show the uniformity of variation. While only P-state resistance variation is shown here, AP-state resistance shows similar levels of variation. In 100 cycles, the resistance of the MTJ varies less than 5~\%, suggesting this variation has a trivial contribution to any variation in performance of inference.

\begin{figure*}
        \centering
        \includegraphics[width=0.9\linewidth]{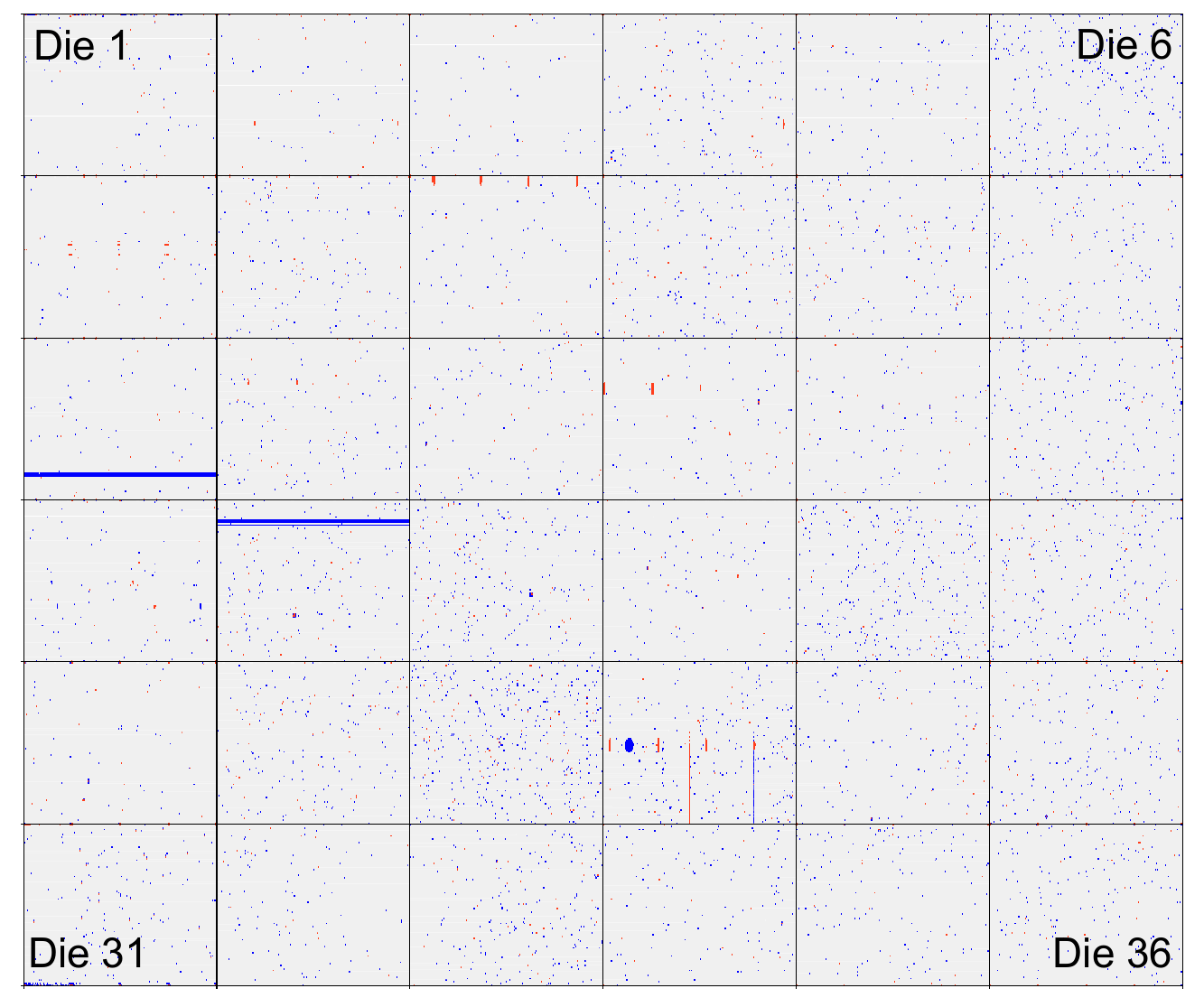}\par
        \caption{Defect maps of the 36 dies used in this work. Axis labels have been removed for ease of viewing. Each die has been filtered to show locations of electrically shorted devices (red) and subpar devices (blue). }
        \label{fig:yield}
\end{figure*}

Fig.~\ref{fig:yield} plots the 36 dies used in this work, showing electrically shorted locations (red) and subpar devices (blue). A large number of defects can be seen in dies 13, 20, and 28. These are due to non-idealities introduced during fabrication that have shorted pads on the outside of the die which are used to source voltage. The large area in die 28 is due to similar effects but located in a central region of the die.

\bibliography{AltComp}
\end{document}